\documentclass[twocolumn,showpacs,preprintnumbers,amsmath,amssymb]{revtex4}
\usepackage{amsfonts,amssymb,latexsym,amsmath,amscd}
\usepackage{times,mathptm,color}

\newcommand{\RR}{\mathbb{R}}

\newcommand{\ff}{\gamma}

\providecommand{\implies}{\Rightarrow}

    {
    \smallskip
    \refstepcounter{theorem}
    \noindent
    {\bf Example \Roman{section}.\arabic{theorem}} \ \ }
    {\hspace*{\fill}{$\Diamond$}
                  \smallskip}

    {
    \smallskip
    \refstepcounter{theorem}
    \noindent
    {\bf Remark \Roman{section}.\arabic{theorem}} \ \ }
    {\hspace*{\fill}{$\Diamond$}
    \smallskip}

\newenvironment{definition}
    {
    \smallskip
    \refstepcounter{theorem}
    \noindent
    {\bf Definition \Roman{section}.\arabic{theorem}} \ \ }
    {\hspace*{\fill}{\ }
    \smallskip}

\newenvironment{proof}[1][]
    {
    \noindent
    {\bf Proof{#1}:  }
    }
    {\hspace*{\fill}{$\Box$}\smallskip}

    {
    \noindent
    {\bf Sketch{#1}:  }
    }
    {\hspace*{\fill}{$\Box$}\smallskip}

    {
    \noindent
    {\bf Reference:  }
    }
    {\hspace*{\fill}{$\odot$}\smallskip}

\newtheorem{theorem}{Theorem}[section]
\newtheorem{proposition}[theorem]{Proposition}

\newtheorem{lemma}[theorem]{Lemma}
\newtheorem{corollary}[theorem]{Corollary}

\newcommand{\botCNOT}{
\begin{picture}(0,0)
    \put(1,3){\circle{1}}
    \put(0,1){\line(1,0){2.3}}
    \put(1,1){\line(0,1){2.5}}
    \put(1,1){\circle*{0.4}}
    \put(0,3){\line(1,0){2.3}}
\end{picture}
}

\newcommand{\topCNOT}{
\begin{picture}(0,0)
    \put(1,3){\circle*{0.4}}
    \put(0,3){\line(1,0){2.3}}

    \put(1,1){\circle{1}}
    \put(1,0.5){\line(0,1){2.5}}
    \put(0,1){\line(1,0){2.3}}
\end{picture}
}

\newcommand{\boxGate}[1]{
\begin{picture}(0,0)
    \put(0.3,0.3){\framebox(1.4,1.4){\small #1}}
    \put(0,1){\line(1,0){0.3}}
    \put(1.7,1){\line(1,0){0.3}}
\end{picture}
}

\newcommand{\hWire}{
   \begin{picture}(0,0)
    \put(0,1){\line(1,0){2}}
   \end{picture}
}

\begin{document}


\title{Minimal Universal Two-Qubit {\tt CNOT}-based Circuits}

\author{
\begin{tabular}{ccccc}
Vivek V. Shende$^1$ & \;\;\;\; & Igor L. Markov$^2$ & \;\;\;\;
& Stephen S. Bullock$^3$ \\
\footnotesize \tt vshende@umich.edu & & \footnotesize \tt imarkov@umich.edu &
& \footnotesize \tt stephen.bullock@nist.gov \\
\end{tabular}
}

\affiliation{\small
$^1$ The University of Michigan, Department of Mathematics\\
$^2$ The University of Michigan, Department of Electrical Engineering
and Computer Science \\
$^3$ National Institute of Standards and Technology,
I.T.L.-M.C.S.D. }

\date{\today}

\begin{abstract}
   We give quantum circuits that simulate an arbitrary two-qubit unitary
   operator up to global phase. For several quantum gate libraries 
   we prove that gate counts are optimal in worst and average cases.
   Our lower and upper bounds compare favorably to previously published results.
   Temporary storage is not used because it tends to be expensive in physical
   implementations. For each gate library, best gate counts can be achieved by a
   single universal circuit.
   To compute gate parameters in universal circuits,
   we only use closed-form algebraic expressions,
   and in particular do not rely on matrix exponentials.
   Our algorithm has been coded in C++.
\end{abstract}

\maketitle

\section{Introduction}
\label{sec:intro}

 Recent empirical work on quantum communication, cryptography and computation
 \cite{NielsenC:00}
 resulted in a number of experimental systems that can implement two-qubit
 circuits. Thus, decomposing arbitrary two-qubit operators into
 fewer gates from a universal library may simplify such physical
 implementations. While the universality of various gate libraries
 has been established in the past \cite{DiVincenzo:95,BarencoEtAl:95},
 the minimization of gate counts has only been studied recently.
 Universal quantum circuits with six, four and three {\tt CNOT} gates
 have been found that can simulate an arbitrary two-qubit operator
 up to phase \cite{ZhangEtAl:03,BullockM:03,VidalDawson:03, VatanWilliams:03}.
 It has also been shown that if the {\tt CNOT} gate is the only
 two-qubit gate available, then three {\tt CNOT} gates
 are required \cite{VidalDawson:03, VatanWilliams:03,
 OptCUgates:03}. Many of these results rely on the
 Makhlin invariants \cite{Makhlin:00} or the related
 {\em magic basis} and {\em canonical decomposition}
 \cite{BennettEtAl:96,HillWooters:97,LewensteinEtAl:01, KhanejaBG:01a}.
 Similar invariants have been investigated previously \cite{Rains:97,
 Grassl:98} and more recently in \cite{BullockBrennen:03}.

 Our work improves or broadens each of the above circuit constructions
 and lower bounds, as summarized in Table \ref{tab:gatecounts}. We
 rely on the Makhlin invariants \cite{Makhlin:00}, and simplify them
 for mathematical and computational convenience --- our version
 facilitates circuit synthesis algorithms.
 We have coded the computation of
 specific gate parameters in several hundred lines of C++,
 and note that it involves only closed-form algebraic expressions
 in the matrix elements of the original operator (no matrix logarithms
 or exponents) . We articulate the degrees of freedom in our algorithm,
 and our program produces multiple circuits for the same operator.
 This may be useful with particular implementation technologies where
 certain gate sequences are more likely to experience errors.
 Additionally, this paper contributes a lower bound
 for the number of {\tt CNOT} gates required to simulate an 
 arbitrary $n$-qubit
 operator, which is tighter than the generic bound for arbitrary
 two-qubit operators \cite{BarencoEtAl:95, Knill:95}.

 The two lines in Table \ref{tab:gatecounts} give gate counts
 for circuits consisting of elementary and basic gates, respectively.
 Both types were introduced in \cite{BarencoEtAl:95},
 but basic gates better reflect gate costs in some physical implementations
 where all one-qubit gates are equally accessible. Yet, when working
 with ion traps, $R_z$ gates are significantly easier to implement than
 $R_x$ and $R_y$ gates \cite{WinelandEtAl:98}. Our work uncovers another
 asymmetry, 
 which is of theoretical nature and does not depend on
 the implementation technology --- a subtle complication arises
 when only {\tt CNOT}, $R_x$ and $R_z$ gates are available.

 Our work shows that basic-gate circuits can be simplified by
 temporarily decomposing basic gates into elementary gates, so as to apply
 convenient circuit identities summarized in Table \ref{tab:ident}. Indeed,
 all lower bounds in Table \ref{tab:gatecounts} and the $n$-qubit
 {\tt CNOT} bound above rely on these circuit identities.
 Additionally, temporary decompositions into elementary gates
 may help optimizing pulse sequences in physical implementations.

\vbox{
  The remainder of this paper is structured as follows.
  Section \ref{sec:background} discusses gate libraries and
  circuit topologies.  Section \ref{sec:lowerbounds} derives the lower bounds of
  Table \ref{tab:gatecounts}. Section \ref{sec:invariants}
  classifies two-qubit operators up to local unitaries.
  Section \ref{sec:param} develops some technical lemmata, and
  Section \ref{sec:18} constructs small circuits that match upper bounds in
  Table \ref{tab:gatecounts}. Subtle complications caused by the lack
  of the $R_y$ gate are discussed in the Appendix and Section
  \ref{sec:conclusions}.
}

\begin{table}[t]
  \begin{center}
    \begin{tabular}{|l||r|r|r|r|}
      \hline
      Gate libraries&\multicolumn{4}{c|}{\ \ \ \ {\bf Lower} and {\bf Upper} Bounds}\\
              &  {\tt CNOT} & overall     &  {\tt CNOT}  &   overall  \\
      \hline \hline
      \{{\tt CNOT}, any 2 or 3 of \{$R_x$, $R_y$, $R_z$\}\} & 3 & 18 & 3 & 18 \\
      \{{\tt CNOT}, arbitrary 1-qubit gates \} & 3 &  9 & 3 & 10 \\ \hline
    \end{tabular}
  \end{center}
  \caption{
    \label{tab:gatecounts} Constructive upper bounds on gate
    counts for generic circuits using several gate libraries. Each
    bound given for controlled-not ({\tt CNOT}) gates is compatible
    with the respective overall bound. These bounds are tighter than
    those from \cite{BullockM:03,ZhangEtAl:03} in all relevant cases.
  }
\end{table}

\section{Gate libraries and circuit topologies}
\label{sec:background}

We recall that the Bloch sphere isomorphism \cite{NielsenC:00}
identifies a unit vector $\vec{n} = (n_x, n_y, n_z)$ with
$\sigma_n= n_x \sigma_x + n_y \sigma_y + n_z \sigma_z$. Under this
identification, rotation by the angle $\theta$ around the vector
$\vec{n}$ corresponds to the special unitary operator $R_n(\theta)
= e^{-i\sigma_n\theta/2}$. It is from this identification that the
decomposition of an arbitrary one-qubit gate $U =\mbox{e}^{i\Phi}
R_z(\theta) R_y(\phi) R_z(\psi)$ arises \cite{NielsenC:00}. Of
course, the choice of $y,z$ is arbitrary; one may take any pair of
orthogonal vectors in place of $\vec{y}, \vec{z}$.

\begin{lemma}
\label{lem:rotations} Let $\vec{n},\vec{m} \in \mathbb{R}^3$,
$\vec{n} \perp \vec{m}$, and $U \in SU(2)$. Then one can find
$\theta, \phi$, and  $\psi$ such that $U = R_n(\theta) R_m(\phi)
R_n(\psi)$.
\end{lemma}

In the case of $\vec{n} \perp \vec{m}$, we have $\sigma_n
R_m(\theta) \sigma_n = R_m(-\theta)$ and $R_n(\pi/2) R_m(\phi)
R_n(-\pi/2) = R_p (\phi)$ for $\vec{p} = \vec{m} \times \vec{n}$.
For convenience, we set $S_n = R_n(\pi/2)$; then $S_z$ is the
usual $S$ gate, up to phase. In the sequel, we always take $m,n$
out of $x,y,z$.

We denote by $C^a_b$ the controlled-not ({\tt CNOT}) gate with
control on the $a$-th qubit and target on the $b$-th. We recall
that $R_z$ gates commute past {\tt CNOT}s on the control line and
$R_x$ gates commute past {\tt CNOT}s on the target. Finally, for
mathematical convenience, we multiply the {\tt CNOT} gate by a
global phase $\xi$ such that $\xi^4 = -1$; to represent it as an
element of $SU(4)$.

In this work we distinguish two types of gate libraries for
quantum operators that are universal in the exact sense (compare
to approximate synthesis and the Solovay-Kitaev theorem).
The {\em basic-gate} library
\cite{BarencoEtAl:95} contains the {\tt CNOT}, and all one-qubit
gates. {\em Elementary-gate} libraries also {\tt CNOT} gate and
one-qubit gates, but we additionally require that they contain
only finitely many one-parameter subgroups of $SU(2)$. We call
these {\em elementary-gate} libraries, and Lemma
\ref{lem:rotations} indicates that if such a library includes two
one-parameter subgroups of $SU(2)$ (rotations about around
orthogonal axes) then the library is universal. In the literature,
it is common to make assertions like: $\dim[SU(2^n)] = 4^n - 1$.
Thus if a given gate library contains only gates from
one-parameter families and fully-specified gates such as {\tt
CNOT}, at least $4^n - 1$ one-parameter gates are necessary
\cite{BarencoEtAl:95}, \cite[Theorem 3.4]{Knill:95}. Such
dimension-counting arguments lower-bound the number of
$R_x,R_y,R_z$ gates required in the worst case
\cite{BarencoEtAl:95}.

To formalize dimension-counting arguments, we introduce the
concept of {\em circuit topologies} --- underspecified circuits
that may have {\em placeholders} instead of some gates, only with
the gate type specified.  Before studying a circuit topology, we
must fix a gate library and thus restrict the types of
fully-specified (constant) gates and placeholders. We say that a
fully-specified circuit $\mathcal{C}$ conforms to a circuit
topology $\mathcal{T}$ if $\mathcal{C}$ can be obtained from
$\mathcal{T}$ by specifying values for the variable gates. All
$k$-qubit gates are to be in $SU(2^k)$, i.e., normalized. For an
$n$-qubit circuit topology $\mathcal{T}$, we define
$Q(\mathcal{T}) \subset SU(2^n)$ to be the set of all operators
that can be simulated, up to global phase, by circuits conforming
to $\mathcal{T}$. We say that $\mathcal{T}$ is universal iff
$Q(\mathcal{T}) = SU(2^n)$. In this work, constant gates are {\tt
CNOT}s, and placeholders represent either all one-qubit gates or a
given one-parameter subgroup of $SU(2)$. We label one-qubit gate
placeholders by $a,b,c,\ldots$, and one-parameter placeholders by
$R_{\ast}$ with subscripts $x$, $y$ or $z$.

We also allow for explicit relations between placeholders. For
example, circuits conforming to the one-qubit circuit topology $a
b a^\dag$ must contain three one-qubit gates and the first and
last must be inverse to each other.

Circuit identities such as $R_n(\theta) R_n(\phi) = R_n(\theta +
\phi)$ can be performed at the level of circuit topologies. This
identity indicates that two $R_n$ gates may always be combined
into one $R_n$ gate, hence anywhere we find two consecutive $R_n$
placeholders in a circuit topology $\mathcal{T}$, we may replace
them with a single one without shrinking $Q(\mathcal{T})$. Of
course, $Q(\mathcal{T})$ does not grow, either, since $R_n(\psi) =
R_n(0) R_n(\psi)$. We may similarly conglomerate arbitrary
one-qubit gate placeholders, pass $R_z$ ($R_x$) placeholders
through the control (target) of {\tt CNOT} gates, decompose
arbitrary one-qubit gate placeholders into $R_n R_m R_n$
placeholders for $n \perp m$, etc.

We now formalize the intuition that the dimension of $SU(2^n)$ should
match the number of one parameter gates.

\begin{lemma}
\label{lem:dimcount} Fix a gate library consisting of constant
gates and finitely many one-parameter subgroups. Then almost all
$n$-qubit operators cannot be simulated by a circuit with fewer
than $4^n - 1$ gates from the one-parameter subgroups.
\end{lemma}

\begin{proof}
Fix a circuit topology $\mathcal{T}$ with fewer than $\ell <
4^n-1$ one-parameter placeholders. Observe that matrix
multiplication and tensor product are infinitely differentiable
mappings and let $f:\mathbb{R}^{\ell}\rightarrow SU(2^n)$ be the
smooth function that evaluates the operator simulated by
$\mathcal{T}$ for specific values of parameters in placeholders.
Accounting for global phase, $Q(\mathcal{T}) = \bigcup_{\xi^{2^n}
= 1} \mbox{Image}(\xi f)$. Sard's theorem
\cite[p.39]{GuilleminPollack:74} demands that $\mbox{Image}(\xi
f)$ be a measure-zero subset of $SU(2^n)$ for dimension reasons,
and a finite union of measure-zero sets is measure-zero.

For a given library, there are only countably many circuit
topologies. Each captures a measure-zero set of operators, and
their union is also a measure-zero set.
\end{proof}

\section{Lower bounds}
\label{sec:lowerbounds}

Lemma \ref{lem:dimcount} implies that for any given elementary gate
library, one can find $n$-qubit operators requiring at least $4^n-1$
one-qubit gates. We use this fact to obtain
a lower bound for the number of {\tt
CNOT} gates required.

\begin{proposition}
\label{prop:cnotcount} Fix any gate library containing only the
{\tt CNOT} and one-qubit gates. Then almost all $n$-qubit
operators cannot be simulated by a circuit with fewer than $\lceil
\frac{1}{4}(4^n - 3n -1)\rceil$ {\tt CNOT} gates.
\end{proposition}

\begin{proof} Enlarging the gate library cannot increase
the minimum number of {\tt CNOT}s in a universal circuit. Thus we
may assume the library is the basic-gate library. We show that any
$n$-qubit circuit topology $\mathcal{T}$ with $k$ {\tt CNOT} gates
can always be replaced with an $n$-qubit circuit topology
$\mathcal{T}'$ with gates from the \{$R_z$, $R_x$, {\tt CNOT}\}
gate library such that $Q(\mathcal{T}) = Q(\mathcal{T}')$ and
$\mathcal{T}'$ has $k$ {\tt CNOT}s and at most $3n + 4k$
one-parameter gates. The proposition follows from $3n + 4k \ge 4^n
- 1$.

We begin by conglomerating neighboring one-qubit gates; this
leaves at most $n + 2k$ one-qubit gates in the circuit. Now
observe that the following three circuit topologies parametrise
the same sets of operators:
\[C_1^2 (a \otimes b) = C_1^2 (R_x R_z R_x \otimes R_z R_x R_z) =
(R_x \otimes R_z) C_1^2 (R_z R_x \otimes R_x R_z)\]
We use this
identity iteratively, starting at the left of the circuit
topology. This ensures that each {\tt CNOT} has exactly four
one-parameter gates to its left.
(Note that we apply gates in circuits left to right,
but read formulae for the same circuits right to left.)
The $n$ one-qubit gates at the far right of the circuit
can be decomposed into three one-parameter gates apiece.
\end{proof}

\begin{corollary}
\label{cor:bounds} Fix an elementary-gate library. Then almost all
two-qubit operators 
cannot be simulated without at least
three {\tt CNOT} gates and fifteen one-qubit gates.
\end{corollary}

For elementary-gate libraries containing two out of the three
subgroups $R_x, R_y, R_z$, we give explicit universal two-qubit
circuit topologies matching this bound in Section \ref{sec:18}.

\begin{proposition}
{\label{lem:bounds:basic} Using the basic-gate library, almost all
two-qubit operators require at least three {\tt CNOT} gates, and
at least basic nine gates total.}
\end{proposition}
\begin{proof}
Proposition \ref{prop:cnotcount} implies that at least three {\tt
CNOT} gates are necessary in general; at least five one-qubit
placeholders are required for dimension reasons. The resulting
overall lower bound of eight basic gates can be improved further
by observing that given any placement of five one-qubit gates
around three {\tt CNOT}s, one can find two one-qubit gates on the
same wire, separated only by a {\tt CNOT}. Using the $R_z R_x R_z$
or $R_x R_z R_x$ decomposition as necessary, the 5 one-qubit gates
can be replaced by fifteen one-parameter gates in such a way that
the closest parameterized gates arising from the adjacent
one-qubit gates can be combined. Thus, if five one-qubit
placeholders and three {\tt CNOT}s suffice, then so do fourteen
one-parameter placeholders and three {\tt CNOT}s, which
contradicts dimension-based lower bounds.
\end{proof}

\section{Invariants of two-qubit operators}
\label{sec:invariants}

To study two-qubit operators that differ only by pre- or
post-composing with one-qubit operators, we use the
terminology of {\em cosets}, common in abstract algebra \cite{Artin:91}.
Let $G$ be the group of operators
that can be simulated entirely by one-qubit operations. That is,
$G = SU(2)^{\otimes n} = \{a_1 \otimes a_2 \otimes \ldots \otimes
a_n:a_i \in SU(2)\}$. Then two operators $u, v$ are said to be in
the same left coset of $SU(4)$ modulo $G$ (written: $uG=vG$) iff
$u$ differs from $v$ only by pre-composing with one-qubit
operators; that is, if $u = vg$ for some $g \in G$. Similarly, we
say that $u$ and $v$ are in the same right coset ($Gu = Gv$) if
they differ only by post-composition ($u=hv$ for some $h\in G$),
and we say that $u$ and $v$ are in the same double coset ($u =
GvG$) if they differ by possibly both pre- and post-composition
($u = hvg$ for some $g,h \in G$). In the literature, the double
cosets are often referred to as {\em local equivalence classes}
\cite{ZhangEtAl:03}.

Polynomial invariants classifying the double cosets have been
proposed by Makhlin \cite{Makhlin:00}.
In what follows, we present equivalent invariants which generalize
to $n$-qubits and are more straightforward to compute.
Moreover, the proofs given here detail
an explicit constructive procedure to find $a, b, c, d$ such that
$(a \otimes b) u (c \otimes d) = v$, once it has been determined
by computing invariants that $u, v$ are in the same double coset.

\begin{definition}
  \label{def:f}
  We define $\ff_n$ on $2^n \times 2^n$ matrices by the formula
  $u \mapsto u \sigma_y^{\otimes n} u^T \sigma_y^{\otimes n}$.
  When $n$ is arbitrary or clear from context, we write $\ff$ for $\ff_n$.
\end{definition}

\begin{proposition}
  \label{prop:f}
  $\ff$ has the following properties:\\

  1. $\ff(I) = I$

  2. $\ff(ab) = a \ff(b) \ff(a^T)^T a^{-1}$

  3. $\ff(a \otimes b) = \ff(a) \otimes \ff(b)$

  4. $g \in M_{2 \times 2}^{\otimes n} \implies \ff(g) = \det(g) \cdot I$

  5. $\ff$ is constant on the left cosets $u \cdot SU(2)^{\otimes n}$

  6. $\chi[\ff]$ is constant on double cosets $SU(2)^{\otimes n}
      \cdot u \cdot SU(2)^{\otimes n}$
\end{proposition}

\begin{proof} (1), (2), and (3) are immediate from the definition.
(4) can be checked explicitly for $n=1$, and then the general case
follows from (3). For (5), note first that $g \in SU(2)^{\otimes
n} \implies \ff(g) = I$ by (4). Then expressing $\ff(ag)$ and
$\ff(a\cdot I)$ using (1) and (2), we see they are equal. For (6),
we use (2), (4), and (5) to see that $g,h \in SU(2)^{\otimes n}
\implies \ff(gah) = g^{-1} \ff(ah) g = g^{-1} \ff(a) g$ thus
$\chi[\ff(gah)] = \chi[\ff(a)]$. Incidentally, (6) is closely
related to \cite[Thm I.3]{BullockBrennen:03}.
\end{proof}

While $\ff$ is constant on left cosets and $\chi[\ff]$ on double
cosets, these invariants do not in general suffice to classify
cosets. Roughly, a parameter space for double cosets would need
dimension $\dim(SU(2^n)) - 2 \dim (SU(2)^{\otimes n}) = 4^n - 6n
-1$, whereas the space of possible $\chi[\ff]$ has dimension $2^n
- 1$ (because the $2^n$ roots of $\chi(\ff)$ must all have unit
length and have unit product). The first dimension is much larger
except for $n=1,2$. In the case $n=1$, there is only one left
coset (and only one double coset), so our invariants trivially
suffice. For $n=2$, these numbers come out exactly equal, and
$\gamma$ and $\chi[\gamma]$ serve to classify respectively the
left cosets and double cosets.

\begin{proposition} \label{prop:invariants}
For $u, v \in SU(4)$, $G = SU(2) \otimes SU(2)$: \\

  1. $u \in G \iff \ff(u) = I$

  2. $uG = vG \iff \ff(u) = \ff(v)$

  3. $GuG = GvG \iff \chi[\ff(u)] = \chi[\ff(v)]$

\end{proposition}

\begin{proof}
  Recall that $E \in U(4)$ can be found such that $E~SO(4)~E^\dag =
  G$; such matrices are characterized by the property that
  $EE^T = -\sigma_y \otimes \sigma_y$. This and related issues
  have been exhaustively dealt with in several papers
\cite{BennettEtAl:96,HillWooters:97,LewensteinEtAl:01, 
  KhanejaBG:01a,BullockBrennen:03},
  where it is shown that $E$ can be chosen as:
{
  \[
\frac{1}{\sqrt{2}} \left(\footnotesize
    \begin{array}{cccc}
        1 & i & 0 & 0 \\
        0 & 0 & i & 1 \\
        0 & 0 & i & -1 \\
        1 & -i & 0 & 0 \\
    \end{array}
    \right)
  \]
}
  Observe that the properties $\ff(u) = I, \ff(u)=\ff(v),
  \chi[\ff(u)] = \chi[\ff(v)]$ are not changed by replacing
  $\ff$ with $E^\dag \ff E$. Then using the fact
  $- \sigma_y \otimes \sigma_y = EE^T = (EE^T)^\dag$ compute:
  \[E^\dag \ff(g) E = E^\dag g E E^T g^T E^{t\dag} E^\dag E
  = (E^\dag g E)(E^\dag g E)^T\]
  Therefore it suffices to prove the
  proposition after making the following substitutions: $g \mapsto u=E^\dag
  g E$, $G \mapsto SO(4)$, $\ff(g) \mapsto uu^T$.
  Now (1) is immediate and (2) follows from
  $uu^T = vv^T \iff v^\dag u = (v^\dag u)^{t\dag} \iff v^\dag u \in SO(4)$

  To prove (3), note that for $P$ symmetric unitary, $P^{-1} =
  \overline{P}$, hence $[P+\overline{P}, P-\overline{P}]=0$. It
  follows that the real and imaginary parts of $P$ share an
  orthonormal basis of eigenvectors. As they are moreover real
  symmetric matrices, we know from the spectral theorem that their
  eigenvectors can be taken to be real.
  Thus one can find an $a \in SO(4)$ such
  that $auu^T a^\dag$ is diagonal. By re-ordering (and negating)
  the columns of $a$, we can re-order the diagonal elements of
  $auu^T a^\dag$ as desired. Thus if $\chi[uu^T]=\chi[vv^T]$,
  we can find $a, b \in SO(4)$ such that
  $auu^T a^T = b vv^T b^T$ by diagonalizing both; then
  $(v^\dag b^T a u)(v^\dag b^T a u)^T = I$. Let $c = v^\dag b^T a u \in
  SO(4)$. We have $ a^T b v c = u$, as desired.
\end{proof}
The proof above gives an algorithm for computing $a,b,c,d$ for
given two-qubit $u$ and $v$ so that $(a \otimes b) u (c \otimes d)
= v$. Also, $u$ may be chosen as a relative-phasing of Bell
states.

\section{Technical Lemmata}
\label{sec:param}

We present two parameterizations of the space of double cosets
described in Section \ref{sec:invariants}. These will be used in
the constructions of universal two-qubit circuit topologies to
follow.

We will use the following general technique to compute $\ff(u)$.
First, determine a circuit, $C$, simulating the operator $u$.
Given $C$, it is straightforward to obtain a circuit simulating
$\sigma_y^{\otimes 2} u^T \sigma_y^{\otimes 2}$: reverse the order
of gates in $C$, and replace a given gate $g$ by
$\sigma_y^{\otimes 2} g^T \sigma_y^{\otimes 2}$. As will be shown
below, if $g$ is a one-qubit gate, then $\sigma_y^{\otimes 2} g^T
\sigma_y^{\otimes 2} = g^\dag$. For the {\tt CNOT}, we note that
$\sigma_y^{\otimes 2} C_1^2 \sigma_y^{\otimes 2} = C_1^2 (\sigma_x
\otimes \sigma_z)$ and similarly $\sigma_y^{\otimes 2} C_2^1
\sigma_y^{\otimes 2} = C_2^1 (\sigma_z \otimes \sigma_x)$. Now,
combine the circuits for $u$ and $\sigma_y^{\otimes 2} u^T
\sigma_y^{\otimes 2}$ to obtain a circuit simulating $\ff(u)$.

\begin{proposition} \label{prop:parameters:yz}
For any $u \in SU(4)$, one can find $\alpha, \beta, \delta$
such that $\chi [\ff(u)] = \chi[\ff(C_1^2 (I \otimes R_y(\alpha))
C_2^1 (R_z(\delta) \otimes R_y(\beta)) C_1^2)]$.
\end{proposition}

\begin{proof}
Let $v = C_1^2 (I \otimes R_y(\alpha)) C_2^1 (R_z(\delta) \otimes
R_y(\beta)) C_1^2$. As $v$ is given explicitly by a circuit, we
use the technique described above to determine the following
circuit for $\ff(v)$.

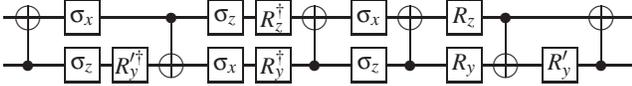
\noindent
\begin{center}
\begin{picture}(26,4)
   \put(0,0){\botCNOT}
   \put(2,0){\boxGate{$\sigma_z$}}
   \put(2,2){\boxGate{$\sigma_x$}}

   \put(4,0){\boxGate{$R_y'^\dag$}}
   \put(4,2){\hWire}

   \put(6,0){\topCNOT}
   \put(8,0){\boxGate{$\sigma_x$}}
   \put(8,2){\boxGate{$\sigma_z$}}
   \put(10,2){\boxGate{$R_z^\dag$}}
   \put(10,0){\boxGate{$R_y^\dag$}}

   \put(12,0){\botCNOT}
   \put(14,0){\boxGate{$\sigma_z$}}
   \put(14,2){\boxGate{$\sigma_x$}}

   \put(16,0){\botCNOT}
   \put(18,0){\boxGate{$R_y$}}
   \put(18,2){\boxGate{$R_z$}}
   \put(20,0){\topCNOT}
   \put(22,0){\boxGate{$R_y'$}}
   \put(22,2){\hWire}
   \put(24,0){\botCNOT}
\end{picture}
\end{center}

Here, $R_y'= R_y(\alpha)$, $R_y = R_y(\beta)$, and $R_z =
R_z(\delta)$. We now use the circuit identities in Figure
\ref{fig:ident} and $\sigma_i R_j(\theta)  = R_j(-\theta)
\sigma_i$ to push all the $\sigma_i$ gates to the left of the
circuit, where they cancel up to an irrelevant global phase of
$-1$. All gates in the wake of their passing become inverted, and
we obtain the following circuit.

\noindent
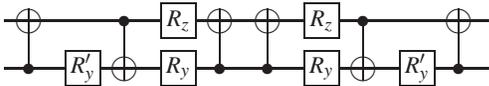
\begin{center}
\begin{picture}(20,4)
   \put(0,0){\botCNOT}
   \put(2,0){\boxGate{$R_y'$}}
   \put(2,2){\hWire}
   \put(4,0){\topCNOT}
   \put(6,2){\boxGate{$R_z$}}
   \put(6,0){\boxGate{$R_y$}}
   \put(8,0){\botCNOT}
   \put(10,0){\botCNOT}
   \put(12,0){\boxGate{$R_y$}}
   \put(12,2){\boxGate{$R_z$}}
   \put(14,0){\topCNOT}
   \put(16,0){\boxGate{$R_y'$}}
   \put(16,2){\hWire}
   \put(18,0){\botCNOT}
\end{picture}
\end{center}

For invertible matrices, $\chi(AB) = \chi(A^{-1}(AB)A) =
\chi(BA)$. In view of the fact that we are ultimately interested
only in $\chi[\ff(V)]$ we may move gates from the left of the
circuit to the right. Thusly conglomerating $R_y'$ gates and
canceling paired {\tt CNOT} gates, we obtain:

\noindent
\begin{center}
\begin{picture}(8,4)
   \put(0,0){\boxGate{$R_y'^2$}}
   \put(0,2){\hWire}
   \put(2,0){\topCNOT}
   \put(4,2){\boxGate{$R_z^2$}}
   \put(4,0){\boxGate{$R_y^2$}}
   \put(6,0){\topCNOT}
\end{picture}
\end{center}

\noindent We have shown $\chi[\ff(v)] = \chi[C_2^1 (R_z(\delta)
\otimes R_y(\beta)) C_2^1 (I \otimes R_y(\alpha))]$. Again, since
$\chi[B] = \chi[A^{-1} B A]$, we conjugate by $I \otimes S_x$.
This fixes the {\tt CNOT} gate and replace $R_y$ gates with $R_z$:
\[\chi[\ff(v)] = \chi[C_2^1(R_z(\delta) \otimes R_z(\beta)) C_2^1 (I
\otimes R_z(\alpha))]\]

Finally, we ensure that the entries of the diagonal matrix
$C_2^1(R_z(\delta) \otimes R_z(\beta)) C_2^1 (I \otimes
R_z(\alpha))$ match the spectrum of $\ff(U)$ by specifying $\alpha
= \frac{x+y}{2}$, $\beta = \frac{x+z}{2}$, and $\delta =
\frac{y+z}{2}$ for $e^{ix}, e^{iy}, e^{iz}$ any three eigenvalues
of $\ff(U)$.
\end{proof}

\noindent
\begin{figure}
    \begin{center}
    \begin{picture}(24,4)
        \put(0,0){\topCNOT}
        \put(2,2){\boxGate{$\sigma_x$}}
        \put(2,0){\hWire}
        \put(5,2){$\equiv$}
        \put(6,0){\boxGate{$\sigma_x$}}
        \put(6,2){\boxGate{$\sigma_x$}}
        \put(8,0){\topCNOT}

        \put(14,0){\botCNOT}
        \put(16,2){\boxGate{$\sigma_z$}}
        \put(16,0){\hWire}
        \put(19,2){$\equiv$}
        \put(20,0){\boxGate{$\sigma_z$}}
        \put(20,2){\boxGate{$\sigma_z$}}
        \put(22,0){\botCNOT}
        \end{picture}
    \end{center}
\caption{\label{fig:ident} Circuit identities to move $\sigma_x$,
$\sigma_z$ past {\tt CNOT}. The $\sigma_x$ identity is standard in
the theory of classical reversible circuits, where $\sigma_x$ is
just the {\tt NOT} gate, and amounts to the statement that $(1
\oplus a) \oplus (1 \oplus b) = (a \oplus b)$. The $\sigma_z$
identity can be obtained from it by conjugating by $H \otimes H$.}
\end{figure}

\begin{proposition} \label{prop:parameters:xz}
For any $u \in SU(4)$, one can find $\theta, \phi, \psi$ such that
$\chi [\ff(u C_2^1 (I \otimes R_z(\psi)) C_2^1)] = \chi[\ff(C_2^1
(R_x(\theta) \otimes R_z(\phi)) C_2^1)]$.
\end{proposition}
\begin{proof}
We set $\Delta = C_2^1 (I \otimes R_z(\psi)) C_2^1$ and compute
$\mbox{tr}[\ff(u\Delta)]$. By Proposition \ref{prop:f}, this is
$\mbox{tr}[\ff(u^T)^T \ff(\Delta)]$. Explicit computation as in
the previous proposition gives $ \ff(\Delta) = \Delta^2$, and one
obtains $\mbox{tr}[\ff(u\Delta)] = (t_1 + t_4)e^{-i \psi} + (t_2 +
t_3)e^{i \psi}$, where $t_1, t_2, t_3, t_4$ are the diagonal
entries of $\ff(u^T)^T$. We may ensure that this number is real by
requiring $\tan(\psi) = \frac{\mbox{Im}(t_1 + t_2 + t_3 +
t_4)}{\mbox{Re}(t_1 + t_2 - t_3 - t_4)}$.

Now consider $m \in SU(N)$, $\chi[m] = \sum a_i X^i = \prod(X -
r_i)$, where the $r_i$ form the spectrum of $m$. Since $m \in
SU(N)$, we must have $\prod r_i = 1 = \prod \overline{r_i}$.
Therefore, $\chi[m] = \chi[m] \prod \overline{r_i} =
\prod(\overline{r_i} X - 1)$. Expanding the equality $\prod(X -
r_i) = \prod(\overline{r_i} X - 1)$ gives $\overline{a_i} =
a_{N-i}$. In particular, for $N = 4$, $a_2 \in \RR$, and
$\mbox{tr}(m) = a_3 = \overline{a_1}$. Since $a_4 = a_0 = 1$,
$\chi[m]$ has all real coefficients iff $\mbox{tr}[m] \in \RR$. In
this case, the roots of $\chi[m]$ must come in conjugate pairs:
$\chi(m) = (X - e^{i r})(X - e^{-i r})(X - e^{is})(X - e^{-is})$.
On the other hand, for $w = C_2^1 (R_x(\frac{r+s}{2})\otimes
R_z(\frac{r-s}{2})) C_2^1$, one can verify that $\chi[\ff(w)]$
takes this form.

Taking $m = \gamma(U C_2^1 (I \otimes R_z(\psi)) C_2^1)$, with
$\psi$ as determined above, we obtain $\theta = \frac{r+s}{2}$,
$\phi = \frac{r-s}{2}$.
\end{proof}

\section{Minimal two-qubit circuits}
\label{sec:18}

We now construct universal two-qubit circuit topologies that match
the upper bounds of Table \ref{tab:gatecounts}. We consider three
different gate libraries: each contains the {\tt CNOT}, and two
out of the three one-parameter gates \{$R_x$, $R_y$, $R_z$\}. We
will refer to these as the CXY, CYZ, and CXZ gate libraries.

In view of Lemma \ref{lem:rotations}, one might think that there
is no significant distinction between these cases. Indeed,
conjugation by the Hadamard gate transforms will allow us to move
easily between the CXY and CYZ gate libraries. However, we will
see that the CXZ gate library is fundamentally different from the
other two. Roughly, the reason is that $R_x$ and $R_z$ can be
respectively moved past the target and control of the {\tt CNOT}
gate, while no such identity holds for the $R_y$ gate. While the
CXY and CYZ libraries each only contain one of \{$R_x$, $R_z$\},
the CXZ gate library contains both, and consequently has different
characteristics. Nonetheless, gate counts will be the same in all
cases. We begin with the CYZ case, which has been previously
considered in \cite{BullockM:03}.

\begin{theorem}
\label{thm:18:cyz} Fifteen \{$R_y$, $R_z$\} gates and three {\tt
CNOT}s suffice to simulate an arbitrary two-qubit operator.
\end{theorem}

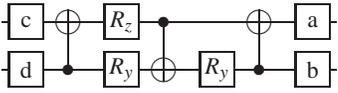
\begin{figure}[t]
\begin{center}
\begin{picture}(14,4)
   \put(0,0){\boxGate{d}}
   \put(0,2){\boxGate{c}}
   \put(2,0){\botCNOT}
   \put(4,0){\boxGate{$R_y$}}
   \put(4,2){\boxGate{$R_z$}}
   \put(6,0){\topCNOT}
   \put(8,0){\boxGate{$R_y$}}
   \put(8,2){\hWire}
   \put(10,0){\botCNOT}
   \put(12,0){\boxGate{b}}
   \put(12,2){\boxGate{a}}
\end{picture}
\caption{\label{fig:18}
  A universal two-qubit circuit with {\bf three} {\tt CNOT} gates.
  It requires {\bf 10} basic gates \cite{BarencoEtAl:95} or
  {\bf 18} gates from  $\{${\tt CNOT}, $R_y$, $R_z\}$.
}
\end{center}
\end{figure}

\begin{proof}
Choose $\alpha, \beta, \delta$ as in Proposition
\ref{prop:parameters:yz}. Then by Proposition
\ref{prop:invariants}, one can find $a,b,c,d \in SU(2)$ such that
\[U = (a \otimes b)C_1^2 (I \otimes R_y(\alpha)) C_2^1
(R_z(\delta) \otimes R_y(\beta)) C_1^2 (c \otimes d)\] Thus, the
circuit topology depicted in Figure \ref{fig:18} is universal.
\end{proof}

\begin{theorem}
\label{thm:18:cxy} Fifteen \{$R_x$, $R_y$\} gates and three {\tt
CNOT}s suffice to simulate an arbitrary two-qubit operator.
\end{theorem}
\begin{proof}
Conjugation by $H^{\otimes n}$ fixes $SU(2^n)$ and $R_y$. It also
flips {\tt CNOT} gates ($H^{\otimes 2} C_1^2 H^{\otimes 2} =
C_2^1$) and swaps $R_x$ with $R_z$.
\end{proof}

Unfortunately, no such trick transforms CYZ into CXZ. Any such
transformation would yield a universal two-qubit circuit topology
in the CXZ library in which only three one-parameter gates occur
in the middle. We show in the Appendix that no such circuit can be
universal and articulate the implications of this distinction
in Section \ref{sec:conclusions}.
Nonetheless, we demonstrate here a universal two-qubit circuit topology
with gates from the \{$R_x$, $R_z$, {\tt CNOT}\} gate library that contains
$15$ one-qubit gates and $3$ {\tt CNOT} gates.

\begin{figure}[b]
\begin{center}
\begin{picture}(14,4)
   \put(0,0){\boxGate{$R_z$}}
   \put(0,2){\hWire}
   \put(2,0){\topCNOT}
   \put(4,0){\boxGate{d}}
   \put(4,2){\boxGate{c}}
   \put(6,0){\topCNOT}
   \put(8,0){\boxGate{$R_z$}}
   \put(8,2){\boxGate{$R_x$}}
   \put(10,0){\topCNOT}
   \put(12,0){\boxGate{b}}
   \put(12,2){\boxGate{a}}
\end{picture}
\caption{\label{fig:18:cxz} Another universal two-qubit circuit
with {\bf three} {\tt CNOT} gates.
  It requires {\bf 10} basic gates \cite{BarencoEtAl:95} or
  {\bf 18} gates from $\{${\tt CNOT}, $R_x$, $R_z\}$.
}
\end{center}
\end{figure}
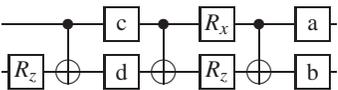

\begin{theorem}
\label{thm:18:cxz}Fifteen \{$R_x$, $R_z$\} gates and three {\tt
CNOT}s suffice to simulate an arbitrary two-qubit operator.
\end{theorem}

\begin{proof}
Let $U'$ be the desired operator; set $U = U' C_2^1 $. Choose
$\theta, \phi, \psi$ for $U'$ as in Proposition
\ref{prop:parameters:xz}. By Proposition \ref{prop:invariants},
one can find $a,b,c,d \in SU(2)$ such that
\[U (I \otimes R_z(\psi)) C_2^1
= (a \otimes b)C_2^1 (R_z(\theta) \otimes R_x(\phi)) C_2^1 (c
\otimes d)\] Solving for $U$ gives the overall circuit topology in
Figure \ref{fig:18:cxz}.
\end{proof}

Unlike the circuit of \ref{thm:18:cyz}, the circuit in Figure
\ref{fig:18:cxz} can be adapted to both other gate libraries. We
can replace $c$ by $S_z ( S_z^\dag c)$ and $a$ by $(a S_z)
S_z^\dag$, then use the $S_z, S_z^\dag$ gates to change the $R_x$
gate into an $R_z$. A similar trick using $R_x$ can change the
bottom $R_z$ gates into $R_y$; this yields a circuit in the CYZ
gate library. As in Theorem \ref{thm:18:cxy}, conjugating by $H
\otimes H$ yields a circuit in the CXY gate library.


Given an arbitrary two-qubit operator, individual gates in
universal circuits can be computed by interpreting proofs of
Propositions \ref{prop:parameters:xz}, \ref{prop:parameters:yz},
and \ref{prop:invariants}, Theorems \ref{thm:18:cyz},
\ref{thm:18:cxy} and \ref{thm:18:cxz} as algorithms. By
re-ordering eigenvalues in the proof of Proposition
\ref{prop:invariants}, one may typically produce several different
circuits. Similar degrees of freedom are discussed in
\cite{BullockM:03}.

To complete Table \ref{tab:gatecounts}, count {\em basic} gates in
Figure \ref{fig:18} or \ref{fig:18:cxz}.

\section{Conclusions}
\label{sec:conclusions}

   Two-qubit circuit synthesis is relevant to on-going physics experiments
   and can be used in peephole optimization of larger circuits, where
   small sub-circuits are identified and simplified one at a time.
   This is particularly relevant to quantum communication, where protocols
   often transmit one qubit at a time and use encoding/decoding circuits
   on three qubits.

   We constructively synthesize small circuits for arbitrary two-qubit operators
   with respect to several gate libraries. Most of our lower and upper bounds
   on worst-case gate counts are tight, and rely on circuit identities
   summarized in Table \ref{tab:ident}.
   We also prove that $n$-qubit circuits
   require $\lceil \frac{1}{4}(4^n - 3n -1)\rceil$
   {\tt CNOT} gates in the worst case.

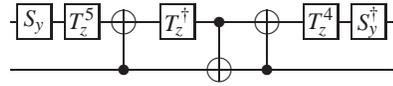
\begin{figure}
        \begin{center}
        \begin{picture}(16,4)
            \put(0,0){\hWire}
            \put(0,2){\boxGate{$S_y$}}

            \put(2,0){\hWire}
            \put(2,2){\boxGate{$T_z^5$}}

            \put(4,0){\botCNOT}

            \put(6,0){\hWire}
            \put(6,2){\boxGate{$T_z^\dag$}}

            \put(8,0){\topCNOT}
            \put(10,0){\botCNOT}
            \put(12,0){\hWire}
            \put(12,2){\boxGate{$T_z^4$}}
            \put(14,0){\hWire}
            \put(14,2){\boxGate{$S_y^\dag$}}
        \end{picture}
        \end{center}
        \caption{\label{fig:qft} The result of our algorithm applied to
         the two-qubit
         Quantum Fourier Transform. The circuit contains 3 one-qubit gates
         and 3 {\tt CNOT}s, but the one-qubit gates are broken up into
     elementary gates
         for specificity. Here, $T_z = R_z(\pi/4)$ is the $T$ gate defined
         in \cite{NielsenC:00} up to a global phase.}
    \end{figure}

   While our techniques do not guarantee
   optimal circuits for non-worst-case operators,
   they perform well in practice: one run of our
   algorithm produced the circuit shown in Figure \ref{fig:qft} for
   the two-qubit Quantum Fourier Transform. We show elsewhere
   that this circuit has minimal basic-gate count.

    A somewhat surprising result of our work is the apparent asymmetry
    between $R_x$, $R_y$ and $R_z$ gates. While one would expect any
    circuit topology for {\tt CNOT}, $R_z$ and $R_y$ to carry over to
    other elementary-gate libraries, we prove a negative result for
    the library {\tt CNOT}, $R_z$ and $R_x$. Namely, using $R_y$ gates
    appears essential for the minimal universal circuit topology shown
    in Figure \ref{fig:18}, which exhibits the maximal possible number
    of one-qubit gates that are not between any two {\tt CNOT} gates.

    The asymmetry between elementary one-qubit gates directly impacts
    peephole optimization of $n$-qubit circuits, where decompositions
    like that in Figure \ref{fig:18} are preferrable over that in
    Figure \ref{fig:18:cxz}. For example, consider a three-qubit
    circuit consisting of two two-qubit blocks on lines (i) one and two,
    (ii) two and three. If both blocks are decomposed as in Figure
    \ref{fig:18}, then the $b$ gate from the first block and the
    $c$ gate from the second block merge into one gate on line two.
    However, no such reduction would happen if the decomposition
    from Figure \ref{fig:18:cxz} is used.

  {
   {\bf Acknowledgments and disclaimers.}  This work is funded by the
   DARPA QuIST program and an NSF grant.
   The views and conclusions contained herein are those of the authors
   and should not be interpreted as neces\-sarily representing official
   policies or endorsements of employers and funding agencies. }

\begin{table}
\begin{center}
\begin{tabular}{|l|l|} \hline
Circuit identities & Descriptions \\
\hline
\hline
   $C^k_j C^k_j = 1$ &
   {\tt CNOT}-gate cancellation\\
   $\omega^{j,k}\omega^{j,k} = 1$ &
   {\tt SWAP}-gate cancellation\\
   $C_j^k C_k^j = \omega^{j,k} C_j^k$ &
   {\tt CNOT}-gate elimination \\
\hline
\hline
   $C^j_k R^j_x(\theta) = R^j_x(\theta) C^j_k$,
   $C^j_k S^j_x = S^j_x C^j_k$ &
   moving $R_x$, $S_x$ via {\tt CNOT} target \\
%
   $C^j_k R^k_z(\theta) = R^k_z(\theta) C^j_k$,
   $C^j_k S^k_z = S^k_z C^j_k$ &
   moving $R_z$, $S_z$ via {\tt CNOT} control \\
\hline

  $\sigma_x^k C_j^k = C_j^k \sigma_x^j \sigma_x^k$ &
  moving $\sigma_x$ via {\tt CNOT} control \\
   $C_j^k\sigma_z^j = \sigma_z^j \sigma_z^k C_j^k$ &
   moving $\sigma_z$ via {\tt CNOT} target \\
\hline
   $C_j^k \omega^{j,k} = \omega^{j,k} C_k^j$ &
   moving {\tt CNOT} via {\tt SWAP} \\
   $V^j \omega^{j,k} = \omega^{j,k} V^k$ &
   moving a 1-qubit gate via {\tt SWAP}  \\

\hline
\hline
   $R_n(\theta) R_n(\phi) = R_n(\theta+\phi)$ &
   merging $R_n$ gates. \\
   $\vec{n} \perp \vec{m} \implies S_n R_m(\theta)=
   R_{n \times m} (\theta) S_n$
   & changing axis of rotation \\
\hline
\end{tabular}
\end{center}
    \caption{\label{tab:ident} Circuit identities used in out work. Here
    $V^j$ represents an arbitrary one-qubit operator acting on wire $j$.}
\end{table}
\section*{Appendix}

We now illustrate the counterintuitive difference between
(i) the CXZ library, and (ii) libraries CYZ and CXY. Namely,
universal circuit topologies with certain properties exist
only for the CYZ and CXY libraries.

The proof of Proposition \ref{thm:18:cyz} contains
a universal generic circuit with three {\tt CNOT} gates and 15 $R_y$ or $R_z$
gates with the property that all but three of the one-qubit gates
appear either before the first or after the last {\tt CNOT} gate.
This is minimal.

\begin{proposition} \label{prop:mid3}
Fix an elementary-gate library. There exist unitary operators $U
\in SU(4)$ that cannot be simulated by any two-qubit circuit in
which all but two of the one-qubit gates appear either before the
first or after the last {\tt CNOT} gate.
\end{proposition}

\begin{proof}
There are four places where the one-parameter gates can appear: at
the left or right of the first or second line. If more than three gates
appear in one such place, conglomerate them into a single one-qubit
gate, and decompose the result into three one-parameter
gates via Lemma \ref{lem:rotations}. By this method, any two-qubit
circuit can be transformed into an equivalent circuit with
at most 12 one-parameter gates on its sides. By Corollary
\ref{cor:bounds}, there exist operators that cannot be simulated
without 15 one-parameter gates; the remaining three must go in the
middle of the circuit.
\end{proof}

We have seen that for the CYZ and the CXY gate libraries, this
lower bound is tight. We will show that this is not the case for
the CXZ gate library. Before beginning the proof, we make several
observations about the CXZ gate library.

Note that conjugating a circuit identity by $H \otimes H$
exchanges $R_x$ and $R_z$ gates, and flips {\tt CNOT}s. Two other
ways to produce new identities from old are: swapping wires, and
inverting the circuit -- reversing the order of gates \& replacing
each with its inverse. For example, one may obtain one of the
commutativity rules below from the other by conjugating by $H
\otimes H$ and then swapping wires.

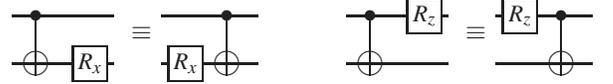
\begin{center}
    \begin{picture}(24,4)
        \put(0,0){\topCNOT}
        \put(2,0){\boxGate{$R_x$}}
        \put(2,2){\hWire}
        \put(5,2){$\equiv$}
        \put(6,0){\boxGate{$R_x$}}
        \put(6,2){\hWire}
        \put(8,0){\topCNOT}

        \put(14,0){\topCNOT}
        \put(16,2){\boxGate{$R_z$}}
        \put(16,0){\hWire}
        \put(19,2){$\equiv$}
        \put(20,0){\hWire}
        \put(20,2){\boxGate{$R_z$}}
        \put(22,0){\topCNOT}
    \end{picture}
\end{center}

When one {\tt CNOT} gate occurs immediately after another in a
circuit, we say that they are {\em adjacent}. When such
pairs of {\tt CNOT}s share control lines, they cancel out, and
otherwise may still lead to reductions as discussed
below.  We will be interested in circuits which do not allow
such simplifications. To this end, recall that $R_x$ gates
commute past the target of a {\tt CNOT}, and $R_z$ gates commute
past the control. Moreover, we have the following circuit
identity: $C_2^1 (R_x(\alpha) \otimes R_z(\beta)) C_2^1 = C_1^2
(R_z(\beta) \otimes R_x(\alpha)) C_1^2$. We say that a given
collection of one-qubit gates {\em effectively separates} a chain
of {\tt CNOT}s iff there is no way of applying the aforementioned
transformation rules to force two {\tt CNOT} gates to be adjacent.
For example, there is no way to effectively separate two {\tt
CNOT}s of opposite orientation by a single $R_x$ or $R_z$
gate. This is illustrated below.

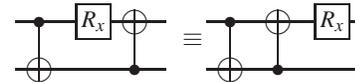
\begin{center}
\begin{picture}(14,4)
    \put(0,0){\topCNOT}
    \put(2,0){\hWire}
    \put(2,2){\boxGate{$R_x$}}
    \put(4,0){\botCNOT}
    \put(7,2){$\equiv$}
    \put(8,0){\topCNOT}
    \put(10,0){\botCNOT}
    \put(12,2){\boxGate{$R_x$}}
    \put(12,0){\hWire}
\end{picture}
\end{center}

On the other hand, two {\tt CNOT} gates of the same orientation
can be effectively separated by a single $R_x$ or $R_z$ gate, as
shown below. Up to swapping wires, these are the only ways to
effectively separate two {\tt CNOT}s with a single $R_x$ or $R_z$.

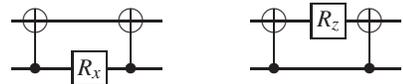
\begin{center}
\begin{picture}(16,4)
    \put(0,0){\botCNOT}
    \put(2,2){\hWire}
    \put(2,0){\boxGate{$R_x$}}
    \put(4,0){\botCNOT}

    \put(10,0){\botCNOT}
    \put(12,2){\boxGate{$R_z$}}
    \put(12,0){\hWire}
    \put(14,0){\botCNOT}
\end{picture}
\end{center}

\begin{proposition} \label{prop:4eff}
At least four gates from \{$R_x$, $R_z$\} are necessary to
effectively separate four or more {\tt CNOT} gates.
\end{proposition}
\begin{proof}
Clearly it suffices to check this in the case of exactly four {\tt
CNOT}s. If three $R_x$, $R_z$ gates sufficed, then one would have
to go between each pair of {\tt CNOT} gates. Suppose all the {\tt
CNOT} gates have the same orientation, say with control on the
bottom wire. Then the first pair must look like one of the pairs
above. In either case, we may use the identity $C_2^1 (R_x(\alpha)
\otimes R_z(\beta)) C_2^1 = C_1^2 (R_z(\beta) \otimes R_x(\alpha))
C_1^2$ to flip these {\tt CNOT} gates, thus ensuring that there is
a consecutive pair of {\tt CNOT} gates with opposite orientations.
As remarked above, there is no way to effectively separate these
using the single one-qubit gate allotted them.
\end{proof}

Denote by $\omega^{ij}$ the {\tt SWAP} gate which exchanges the
$i$-th and $j$-th qubits. It can be simulated using {\tt CNOT}s as
$C_i^j C_j^i C_i^j = \omega^{ij} = C_j^i C_i^j C_j^i$. {\tt SWAP}
gates can be pushed through an elementary-gate circuit without
introducing new gates. So, consider a two-qubit circuit in which
adjacent {\tt CNOT} gates appear. If they have the same
orientation (eg. $C_1^2 C_1^2$ or $C_2^1 C_2^1$), then they cancel
out and can be removed from the circuit. Otherwise, use the
identity $C_1^2 C_2^1 = C_2^1 \omega^{12}$ or $C_2^1 C_1^2 = C_1^2
\omega^{12}$ and push the {\tt SWAP} to the end of the circuit. We
apply this technique at the level of circuit topologies and
observe that since $Q(\mathcal{T}\omega^{12})$ is measure-zero (or
universal) iff $Q(\mathcal{T})$ is. By the above discussion, we
can always reduce to an effectively separated circuit before
checking these properties.

\begin{proposition} \label{prop:mid4}
    Almost all unitary operators $U \in SU(4)$ cannot be simulated
    by any two-qubit circuit with CXZ gates
    in which all but three of the $R_x, R_z$ gates
    appear either before the first or after the last {\tt CNOT}.
\end{proposition}
\begin{proof}
    We show that any circuit topology of the form above can only
    simulate a measure-zero subset of $SU(4)$; the result then
    follows from the fact that a countable union of measure-zero
    sets is measure-zero.

    The assumption amounts to the fact that only three gates are available
    to effectively separate the {\tt CNOT} gates.
    By Proposition \ref{prop:4eff} and the discussion immediately
    following it, we need only consider circuit topologies with no
    more than three {\tt CNOT}s. On the other hand, we know from
    Proposition \ref{lem:bounds:basic} that any two-qubit circuit topology with
    fewer than three {\tt CNOT} gates can simulate only a measure-zero
    subset of $SU(4)$. Thus it suffices to consider circuit
    topologies with exactly three {\tt CNOT} gates. Moreover, we
    can require that they be effectively separated, since
    otherwise we could reduce to a two-{\tt CNOT} circuit.

    Three {\tt CNOT}s partition a minimal two-qubit circuit
    in four regions. We are particularly interested in
    the two regions limited by {\tt CNOTs} on both sides
    because single-qubit gates in those regions must effectively separate
    the {\tt CNOT}s. To this end, we consider
    two pairs of {\tt CNOT}s (the central {\tt CNOT} is in both
    pairs), and distinguish these three cases: (1) both pairs of
    {\tt CNOT}s consist of gates
    of the same orientation, (2) both consist of gates of opposite
    orientations, or (3) one pair has gates of the opposite
    orientations and the other pair has gates of the same
    orientation. In the second case, the {\tt CNOT} gates cannot
    be effectively separated, since each pair of gates with
    opposite orientations requires two one-parameter gates to be
    effectively separated, and only three $R_x$, $R_z$ gates
    are available. In the third case, two {\tt CNOT}s with
    opposite orientations must be separated by two one-parameter
    gates,
    leaving only one $R_x$ or $R_z$ to separate the pair with the same
    orientation. Thus, the pair with the same orientation may be flipped,
    reducing to Case 1, as shown below.

    
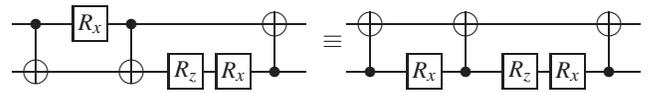
\begin{center}
    \begin{picture}(26,4)
        \put(0,0){\topCNOT}
        \put(2,0){\hWire}
        \put(2,2){\boxGate{$R_x$}}
        \put(4,0){\topCNOT}
        \put(6,2){\hWire}
        \put(8,2){\hWire}
        \put(6,0){\boxGate{$R_z$}}
        \put(8,0){\boxGate{$R_x$}}
        \put(10,0){\botCNOT}
        \put(13,2){$\equiv$}

        \put(14,0){\botCNOT}
        \put(16,2){\hWire}
        \put(16,0){\boxGate{$R_x$}}
        \put(18,0){\botCNOT}
        \put(20,2){\hWire}
        \put(22,2){\hWire}
        \put(22,0){\boxGate{$R_x$}}
        \put(20,0){\boxGate{$R_z$}}
        \put(24,0){\botCNOT}
    \end{picture}
    \end{center}

    Finally, consider the case in which all three {\tt CNOT} gates
    have the same orientation. Each pair of consecutive {\tt
    CNOT}s must have at least one $R_x$ or $R_z$ between them, to
    be effectively separated. Thus one of the pairs has a single $R_x$ or $R_z$
    between its members, and the other has two one-qubit gates.
    We refer to these as the 1-pair and the 2-pair, respectively.

    Suppose that the one-qubit gates separating the 2-pair of
    {\tt CNOT}s occur on different lines. If either one-qubit can commute
    past the {\tt CNOT}s of the 2-pair, then it can move to the edge of the circuit;
    in this case Proposition \ref{prop:mid3} implies that the circuit
    topology we are looking at can only simulate a measure-zero subset of
    $SU(4)$ (one can show that two $R_x$, $R_z$ gates cannot effectively
    separate three {\tt CNOT}s.) Otherwise, we use the identity
    $C_2^1 (R_x(\alpha) \otimes R_z(\beta)) C_2^1 = C_1^2 (R_z(\beta)
    \otimes R_x(\alpha)) C_1^2$ to flip the 2-pair,
    and thus 1-pair now have opposite orientations. As there is
    only one one-qubit gate between them, this pair
    is not effectively separated. For example:

    
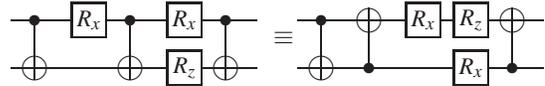
\begin{center}
    \begin{picture}(22,4)
        \put(0,0){\topCNOT}
        \put(2,0){\hWire}
        \put(2,2){\boxGate{$R_x$}}
        \put(4,0){\topCNOT}
        \put(6,0){\boxGate{$R_z$}}
        \put(6,2){\boxGate{$R_x$}}
        \put(8,0){\topCNOT}
        \put(11,2){$\equiv$}
        \put(12,0){\topCNOT}
        \put(14,0){\botCNOT}
        \put(16,0){\hWire}
        \put(16,2){\boxGate{$R_x$}}
        \put(18,0){\boxGate{$R_x$}}
        \put(18,2){\boxGate{$R_z$}}
        \put(20,0){\botCNOT}
    \end{picture}
    \end{center}

    We are left with the possibility that all the {\tt CNOT} gates have
    the same orientation and that the 2-pair's one-qubit gates appear on the
    same line. Both $R_z$, $R_x$ must occur, or else we could combine
    them and apply Proposition \ref{prop:mid3} to show
    that such a circuit topology can only simulate a
    measure-zero subset of $SU(2^n)$. Now, if $R_x R_z$ appears
    between two {\tt CNOT} gates of the same orientation, then
    either the $R_x$ or the $R_z$ can commute past one of them. If
    the outermost gate can commute, Proposition \ref{prop:mid3}
    again implies that the circuit topology simulates only a
    measure-zero subset of $SU(2^n)$. Thus the inner gate can
    commute with the 1-pair. We have now interchanged the roles of
    the 1-pair and the 2-pair, thus by the previous paragraph,
    the gate which originally separated the 1-pair must be on the same line
    as the commuting gate. It follows that all gates are on the same line.
    Up to conjugating by $H \otimes H$, swapping wires, and
    inverting the circuit, this leaves exactly one possibility.

    
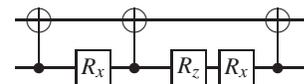
\begin{center}
    \begin{picture}(12,4)
        \put(0,0){\botCNOT}
        \put(2,2){\hWire}
        \put(2,0){\boxGate{$R_x$}}
        \put(4,0){\botCNOT}
        \put(6,2){\hWire}
        \put(8,2){\hWire}
        \put(6,0){\boxGate{$R_z$}}
        \put(8,0){\boxGate{$R_x$}}
        \put(10,0){\botCNOT}
    \end{picture}
    \end{center}

 Finally, we add the four one-qubit gates on the sides, decompose
 each into $R_x R_z R_x$ via Lemma \ref{lem:rotations}, and observe
 that an $R_x$ gate can commute across the top and be absorbed on
 the other side.  This leaves 14
 one-parameter gates, and by Lemma \ref{lem:dimcount}, such a
 circuit topology simulates only a measure-zero subset of $SU(4)$.
\end{proof}

\end{document}